# Terrestrial planet formation from a ring

J. M. Y. Woo[a,b], A. Morbidelli[a], S. L. Grimm[c], J. Stadel[d], R. Brasser[e]

[a]Laboratoire Lagrange, Université Cote d'Azur, CNRS, Observatoire de la Côte d'Azur, Boulevard de l'Observatoire, 06304 Nice Cedex 4, France
[b]Institut für Planetologie, University of Münster, Wilhelm-Klemm-Straße 10, 48149, Germany
[c]Center for Space and Habitability, University of Bern, Gesellschaftsstrasse 6, 3012 Bern, Switzerland
[d]Institute for Computational Science, University of Zürich, Winterthurerstrasse 190, 8057 Zürich, Switzerland
[e]Origins Research Institute, Research Centre for Astronomy and Earth Sciences; Budapest H-1121, Hungary

**Abstract**

It has been long proposed that, if all the terrestrial planets form within a tiny ring of solid material at around 1 AU, the concentrated mass-distance distribution of the current system can be reproduced. Recent planetesimal formation models also support this idea. In this study, we revisit the ring model by performing a number of high-resolution N-body simulations for 10 Myr of a ring of self-interacting planetesimals, with various radial distributions of the gas disc. We found that even if all the planetesimals form at ~1 AU in a minimum mass solar nebula-like disc, the system tends to spread radially as accretion proceeds, resulting in a system of planetary embryos lacking mass-concentration at ~1 AU. Modifying the surface density of the gas disc into a concave shape with a peak at ~1 AU helps to maintain mass concentrated at ~1 AU and solve the radial dispersion problem. We further propose that such a disc should be short lived (≤ 1 Myr) and with a shallower radial gradient in the innermost region (< 1 AU) than previously proposed to prevent a too-rapid growth of Earth. Future studies should extend to ~100 Myr the most promising simulations and address in a self-consistent manner the evolution of the asteroid belt and its role in the formation of the terrestrial planets.

## 1. Introduction

The formation process of the terrestrial planets is still a subject of scientific debate. In general, this process can be divided into two stages. The first stage of terrestrial planet formation describes the process of coagulation of micrometre sized dust until the assembly of km-sized planetesimals (e.g. Drążkowska et al., 2016; Johansen et al., 2014, 2007; Morbidelli et al., 2022; Youdin and Goodman, 2005), while the second stage begins with the mutual collisions of planetesimals to form Mars sized embryos, followed by giant impacts between embryos resulting in a fully formed Earth-sized planet (e.g. Greenberg et al., 1978; Kokubo and Ida, 1996, 1998, 2000, 2002; Leinhardt and Richardson, 2005; Wetherill and Stewart, 1989, 1993). Some fraction of embryos' mass may also be delivered by mm to cm-sized pebbles that drift inward from the outer solar system due to gas drag, although their mass fraction is likely limited (Batygin and Morbidelli, 2022; Lambrechts et al., 2019).The second stage can in turn be divided in two periods: the early phase characterised by the presence of gas in the disc which lasts a few million years and the late phase after gas removal which can last up to 100 Myr. Gas drag on planetesimals and tidal embryo-disc gravitational interactions are dominant processes of the early phase, but totally absent during the late period.

Up to now, none of the current second stage models (Chambers, 2001; Clement et al., 2018; Hansen, 2009; Walsh et al., 2011) has been entirely successful in reproducing both the dynamical characteristics of the planets and their chemical/isotopic properties in relationship with the meteoritic samples. For instance, the Classical model studied in Chambers (2001) constantly forms too massive Mars (e.g. Raymond et al. 2006, 2009). While the Grand Tack model proposed by Walsh et al. (2011) successfully reproduces Mars with the correct mass, the migration of the gas giants induces vigorous radial mixing within the disc, resulting in too similar isotopic composition of Earth and Mars (Woo et al. 2018). The early giant instability model proposed by Clement et al. (2018) has yet to be tested in the isotopic perspective, however the orbits of ice giants after the instability are not well reproduced.

Besides, the previous second-stage models so-far developed have their limitations and thus a revision on these models is required. A first limitation is that the studies of the second stages of planet formation use initial conditions that are not based on a planetesimal formation model. A second limitation is that most studies focused on the post-gas phase of planet formation, which allows using symplectic N-body methods (Chambers, 1999; Duncan et al., 1998; Wisdom and Holman, 1991), but requires making ad-hoc assumptions on the distribution of lunar to Mars-sized embryos on top of asteroid-sized planetesimals (e.g. Brasser et al., 2016; Jacobson and Morbidelli, 2014; Mah and Brasser, 2021; Raymond et al., 2009, 2006). Indeed, embryos should have formed during the gas phase and therefore their formation is not self-consistently simulated in models focusing on the post-gas phase. This work addresses both these limitations.

Concerning planetesimal formation, recent studies suggest that terrestrial planetesimals could form within a narrow region (a few ~0.1 AU) through drifting and piling up of pebbles (Drążkowska et al., 2016). Pebbles could also be trapped in a pressure bump (Haghighipour and Boss, 2003; Izidoro et al., 2022; Masset et al., 2006), creating a local maxima of solid to gas density which enables planetesimals to form via streaming instability (Youdin and Goodman, 2005). Our study follows a recent work which showed that the first planetesimals in the inner solar system should have formed in a narrow ring around the silicate sublimation line, which could be close to ~1 AU (Morbidelli et al., 2022), where most of the current mass of the terrestrial planet system is concentrated. Planetesimals formed at the silicate sublimation line are likely to be volatile poor (Morbidelli et al., 2022), in agreement that the terrestrial planets formed from relatively dry building blocks, isotopically akin to enstatite and ordinary chondrite-like (Brasser et al., 2018; Burkhardt et al., 2021; Dauphas, 2017). In addition, the mass concentration of the terrestrial planets in the Venus-Earth region (~0.7 to 1 AU) independently supports the hypothesis that the terrestrial planets originated from a narrow ring of planetesimals concentrated at ~1 AU (Hansen, 2009; Morishima et al., 2008). Thus, we adopt initial conditions that planetesimals are concentrated in a narrow ring.

Previous N-body simulations have shown that the ring model could potentially explain the mass-distance distribution of the terrestrial planet system (i.e. massive Venus-Earth, low mass Mars and Mercury; Hansen, 2009; Kaib and Cowan, 2015; Lykawka, 2020; Nesvorný et al., 2021; Ogihara et al. 2018). However, one important limitation of these studies is that they assumed that lunar to Mars-sized embryos were distributed tightly within a few ~0.1 AU alongside with small planetesimals. Due to the limitation of computational resources, these studies did not show how embryos emerged from the ring of planetesimals. When studying the ring model, this could pose a potential problem as the ring of solids may not remain narrow during embryo formation due to mutual scattering among objects in the ring. If this occurs, the likelihood of forming a system of planetary embryos concentrated around the current orbits of Venus-Earth could be much lower than assumed in previous studies.

Embryo formation from a population of planetesimals has been modelled in a number of studies, but none of them are suitable for our purpose of simulating terrestrial planet formation from a ring of planetesimals. Indeed several works (Carter et al., 2015; Clement et al., 2020; Walsh and Levison, 2019; Woo et al., 2021) considered planetesimals distributed over a wide disk, following the conventional model termed the "Minimum Mass Solar Nebula" (MMSN, Hayashi, 1981; Weidenschilling, 1977). Embryo formation from a ring of planetesimals was addressed in some studies (Deienno et al., 2019; Lykawka, 2020; Morishima et al., 2008; Walsh and Levison, 2016), but these studies either ignored the presence of a gas disc or did not include the crucial Type-I migration effect (Tanaka et al., 2002; Tanaka and Ward, 2004), which could lead to inward migration of the embryos and cause the ring to spread. We include the gas disc and the effect of Type-I migration in our study.

In this paper, we will study the ring model in a detailed fashion by simulating the early period when embryos form from a ring of real sized planetesimals in the presence of gas. Our gas disc induces gas drag on planetesimals, as well as tidal torque on lunar to Mars size embryos. The major question that we would like to answer is whether (or under which conditions) the ring of terrestrial building material still remains narrow during embryo formation, and results in the mass-concentrated system that we observe today. In the next section, we describe our N-body method and initial conditions adopted. We present our results considering different gas discs in Section 3. In Section 4 we discuss various uncertainties that still exist concerning the ring model. We conclude on our results in the final section.

## 2. Method

### 2.1. GENGA and the "superparticle" model

We adopt the latest version of the GPU N-body code GENGA (Grimm and Stadel, 2014; Grimm et al., 2022) for our study. This GPU accelerated N-body integrator takes advantage of the large number of computing cores in GPU cards which can perform the same instructions on multiple threads in parallel. This speeds up both the mutual force calculation as well as the routines for handling close encounters. The speed increase over a regular single CPU core for $N \sim 8000$ is a factor of 500 (Grimm and Stadel, 2014), where $N$ is the number of gravitationally interacting particles in a simulation. Our tests show that a modern A100 GPU is still an order of magnitude faster than a 96 core AMD EPYC 7552 CPU. Therefore, simulations with a large number ($N > 10,000$) of fully-self gravitating objects can be solved much faster than using conventional CPU-based N-body codes. Several previous studies have adopted GENGA to study terrestrial planet formation starting from a disc of self-interacting planetesimals (Clement et al., 2020; Hoffmann et al., 2017; Woo et al., 2022, 2021).

In this project, we aim to further improve the initial conditions by beginning our simulations with planetesimals of realistic size. According to the theoretical models (Klahr and Schreiber, 2020) and constraints from the current size distribution of asteroids and Kuiper belt objects (Delbo et al., 2017, 2019; Morbidelli et al., 2009; Shankman et al., 2013), the dominant size of the initial planetesimals should have been ~100 km in diameter. Assuming the terrestrial planet system formed from a ring of 100 km planetesimals with a total mass of 2 $M_{Earth}$ requires the system to consist of ~ 8 x $10^6$ self-gravitating particles initially, which exceeds the memory limit even of the latest GPUs. Besides, simulating a system with such a high $N$ would consume too much time in an HPC cluster. To overcome this problem, we have implemented a "super-particle" approach in GENGA.

The super-particle technique is a classic one in N-body codes where planetesimals interact only with proto-planets and not with each other. In these cases, the mass of the planetesimals is enhanced by the factor $N_{ratio}$, which is the ratio between the number of real planetesimals and the number of simulated superparticles in the disc. If gas drag is taken into account, the real masses and sizes of planetesimals are used in the drag force calculation. Our case is more complicated, because planetesimals interact gravitationally with each other. We need to scale masses and sizes in such a way that the real dynamical stirring of the planetesimal population is reproduced, as well as the growth rate. We can achieve this result by adopting different mass-enhancement factors in different parts of the integration algorithm.

The dynamical stirring equation for eccentricity (eq. inclination) of a disc of equal mass particles is proportional to $\rho M$ (Wetherill and Stewart, 1993) where $M$ is the individual mass of the particles and $\rho$ is their spatial density. If we need to reduce the number of simulated particles by $N_{ratio}$ while keeping the same value of $\rho M$, it is clear that the mass of the particles has to be enhanced by $N_{ratio}^{1/2}$. Indeed in this case $\rho$ is decreased by a factor $N_{ratio}^{1/2}$.

Concerning planetesimal growth, it is obvious that, for a given particle, the collision rate is reduced by the $N_{ratio}$ factor if the number of simulated particles is reduced by $N_{ratio}$ and their considered radius is the real one. So, to have the same growth rate d$M$/d$t$ when two particles collide they have to carry a mass that is the real mass enhanced by $N_{ratio}$. The same is true when planetesimals gravitationally interact or collide with proto-planets (here, we consider as a protoplanet any particle that has grown in mass relative to its initial planetesimal mass). Finally, for the calculation of gas drag, both size and mass of planetesimals should be the real ones.

Within this algorithm, the masses of protoplanets is the real one and their size is computed, after each mass-growth episode, by assuming a bulk density of 3 gcm$^{-2}$.

To test the validity of this approach, we have simulated the unrealistic problem of growth of embryos from a ring of $N$ = 7605 planetesimals of individual mass ~2.6 x $10^{-4}$ $M_{Earth}$, which can be treated using the original GENGA algorithm, and compared the results with that of a simulation considering $N/N_{ratio}$ superparticles, with $N_{ratio}$ = 5 (Figure 1 and Figure 2). The agreement is satisfactory for both the growth of embryos (Figure 1) and the dynamical stirring of the disc (Figure 2); the difference in the size of the largest embryos between the two simulations are due to small number statistics and indeed change from test to test.

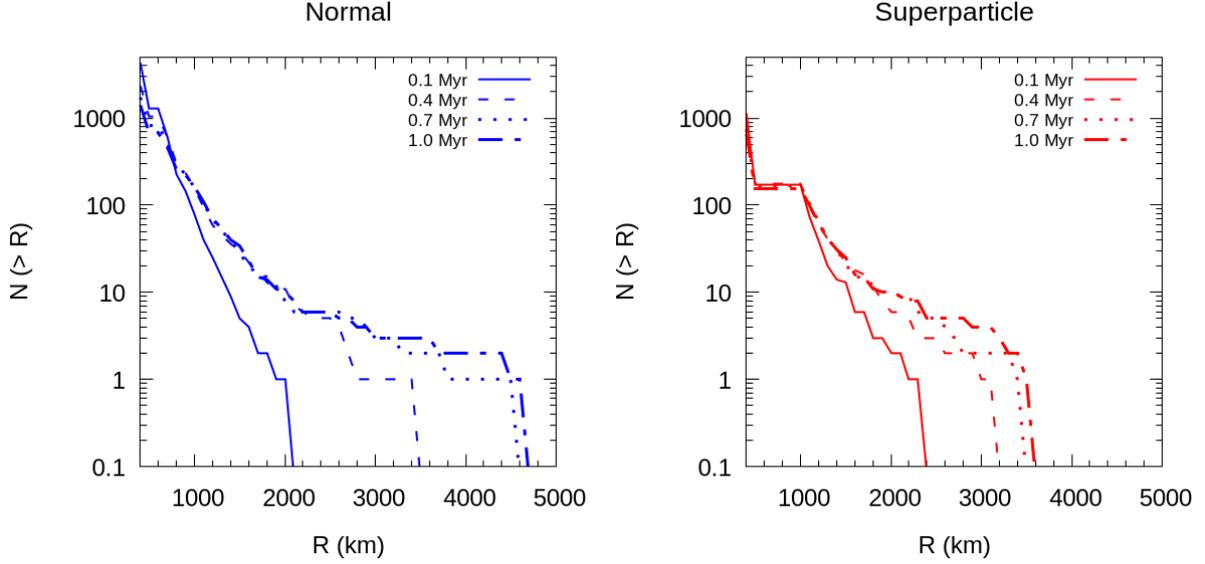

Figure 1 - Comparison of the size-frequency distribution of the normal GENGA simulation (left) with the simulation adopting the super particle method with $N_{ratio} = 5$ (right).

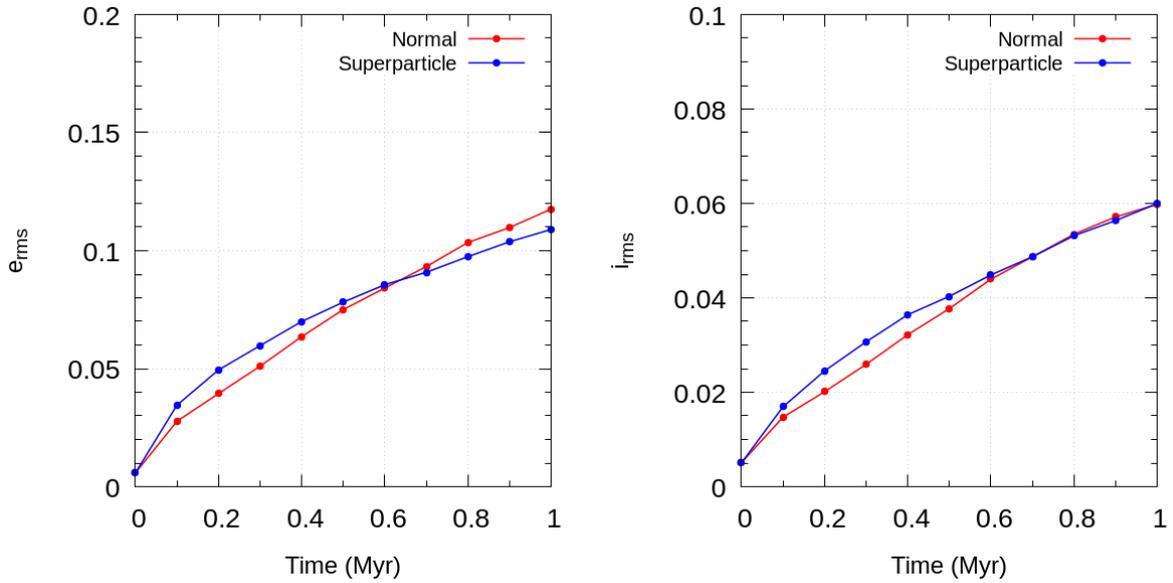

Figure 2 - Comparison of the root mean square of the system's eccentricity (left) and inclination (right) between the normal GENGA simulation (red) and the superparticle simulation with $N_{ratio} = 5$.

In our actual simulations of terrestrial planet formation, we will consider a ring of about 2 Earth masses of 100 km planetesimals, adopting $N_{ratio} = 1000$, which will give us ~8000 particles initially in each simulation. This greatly speeds up the high-resolution simulation at the same time maintaining its physical accuracy.

### 2.2. Initial conditions

We distribute the superparticles within a narrow region according to a gaussian distribution function. Table 1 shows the settings of our initial conditions. The mean location of the ring of superparticles is $\mu = 1$ or, to anticipate some inward migration of the embryos, $\mu = 1.3$ AU. The variance of the distribution is $\sigma = 0.1$ AU for all cases (i.e. 68% of the superparticles are within $\mu \pm 0.1$ AU, and 100 % are within $\mu \pm 0.3$ AU). The total mass of the ring is 2.1 $M_{Earth}$, which is ~10 % more massive than the current system in order to account for loss of mass through mutual scattering. The eccentricities and inclinations of the superparticles are uniformly random in the range $0 < e < 0.01$ and $0° < i < 0.5°$. Their nodal angles and mean anomalies are uniformly random from 0° to 360°. We perform 28 simulations for 10 Myr with all the particles gravitational interacting with each other. Particles closer than 0.1 au or further than 100 au are removed from the simulation.

Table 1 - Initial conditions of our simulations. "Normal gas disc" stands for a disc of gas with density following Eq. (1) while "Convergent disc" describes a modified surface density of the gas disc with a positive radial surface density slope (more or less steep for the nominal vs. shallow case) in the inner region (see Figure 3). $\tau_{decay}$ is the timescale over which the gas density is uniformly reduced.

|  | Gas disc decay timescale, $\tau_{decay}$ | Mean location of ring, $\mu$ | Half width of the ring, $\sigma$ | Total mass of the ring, $M_{tot}$ | Number of superparticles, $N_{super}$ | Number of simulations, $N_{sim}$ |
|---|---|---|---|---|---|---|
| Normal disc | 0.5 Myr | 1 AU | 0.1 AU | 2.1 $M_{Earth}$ | 7983 | 2 |
|  |  | 1.3 AU |  |  |  | 2 |
|  | 1 Myr | 1 AU |  |  |  | 2 |
|  |  | 1.3 AU |  |  |  | 2 |
|  | 2 Myr | 1 AU |  |  |  | 2 |
|  |  | 1.3 AU |  |  |  | 2 |
| Nominal convergent disc | 1 Myr | 1 AU | 0.1 AU | 2.1 $M_{Earth}$ | 7983 | 2 |
|  |  | 1.3 AU |  |  |  | 2 |
|  | 2 Myr | 1 AU |  |  |  | 2 |
|  |  | 1.3 AU |  |  |  | 2 |
| Convergent shallow inner disc | 1 Myr | 1 AU | 0.1 AU | 2.1 $M_{Earth}$ | 7983 | 2 |
|  |  | 1.3 AU |  |  |  | 2 |

|  | 2 Myr | 1 AU |  |  |  | 2 |
|  |  | 1.3 AU |  |  |  | 2 |

The normal gas disc we adopted follows the MMSN with a gas surface density at time *t* is

$$\sum_{gas}(r,t) = \sum_{gas,0}(r/1\ \text{AU})^{-p}\ exp\left(-t/\tau_{\text{decay}}\right),$$

(1)

where *r* is the heliocentric distance from the Sun, $\tau_{\text{decay}}$ = 1 or 2 Myr is the dissipation timescale of the disc, $\Sigma_{gas,0}$ is assumed to be 1600 gcm$^{-2}$ and *p* = 1 (Morishima et al., 2010). The gas disc dissipates globally over time.

We also study another case with a modified gas disc with a gas surface density following Eq. (1) but with *p* = - 2 + 1.25(1 - tanh( (1 - *r*)/0.15) ) (Brož et al., 2021; Haghighipour and Boss, 2003; Lyra et al., 2010; Masset et al., 2006; Ogihara et al., 2018; Zhang et al., 2014), which is compared in Figure 3 to that of the normal disc. Such a modified surface density could be caused by the effect of a magnetic-driven disc wind (Suzuki et al., 2016), which leads to mass loss in the innermost region of the disc. It could also be due to an enhancement of viscosity in the inner part of the disc (Brož et al., 2021). We refer to the modified gas disc as "nominal convergent disc" in our following discussion because the positive surface density slope enhances a positive corotation torque on the protoplanets, leading to their outward migration, whereas planet migration remains inwards where the gas surface density slope is negative. In addition, in order to reduce the strength of convergent migration, we also test a convergent disc with a shallower inner region following the surface density profile as Eq. (1), but with *p* = - 0.5 + 0.5(1 - tanh( (1 - *r*)/0.15) ) (see Figure 3).

We would like to emphasize that, even though we assumed a pressure bump existed at 1 AU, planetesimal formation does not necessarily require such a pressure bump. The accumulation of dust and pebbles at ~ 1 AU can be due to sublimation and recondensation of silicates and to the outward radial flow of the dust in the early phases of the disc (Morbidelli et al. 2022), and this process does not depends strongly on the gas disc profile. Hence, we explore the planetesimals ring profile and gas disc profile in an uncorrelated manner in this study.

We modify the original GENGA script to change the gas surface density, and add the corotational torque into the migration algorithm. All the migration equations follow those descripted by Ogihara et al. (2018) and references therein.

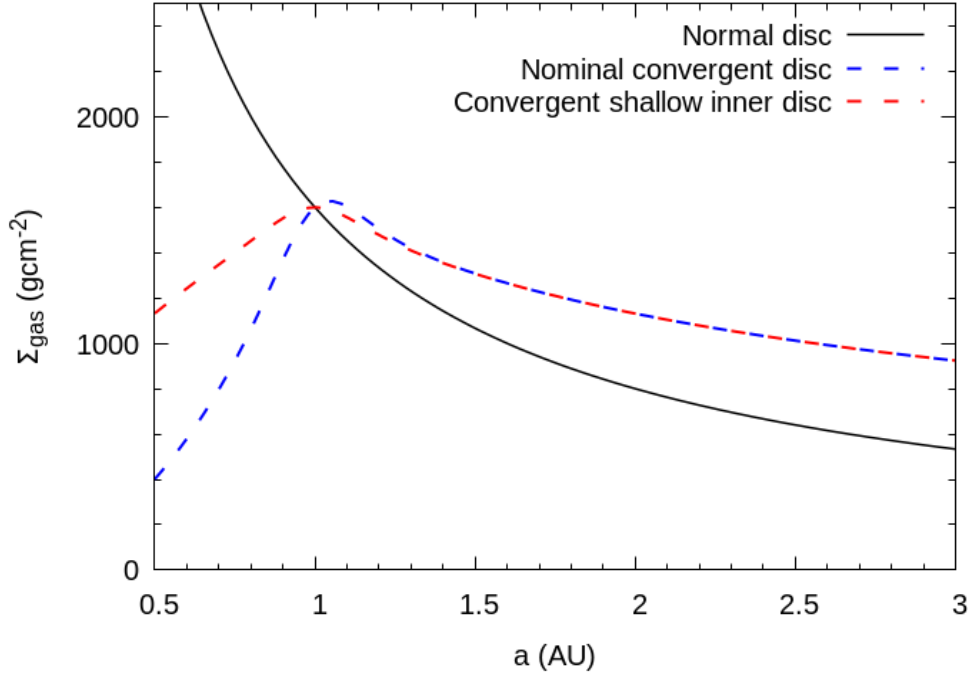

Figure 3 - Initial surface density profile of different gas discs in our study.

Jupiter and Saturn are placed at their suggested pre-instability orbits with more circular and co-planar orbits initially (Tsiganis et al., 2005) (i.e. *Circular Jupiter-Saturn*, CJS: $a_j$ = 5.45 AU, $e_j$ = 0.0009, $i_j$ = 0°; $a_s$ = 8.18 AU, $e_s$ = 0.0002, $i_s$ = 0.5°; where ($a_j, e_j, i_j$) and ($a_s, e_s, i_s$) are the semi-major axis, eccentricity and inclination of Jupiter and Saturn, respectively).

## 3. Results

As we said in the Introduction, we focus our study on the growth of planetary embryos during the gas-disc phase. Thus, our simulations cover only 10 Myr, with an exponentially decaying gas surface density (see Eq. 1). We compare the resulting distribution of embryos with those assumed as initial conditions in successful simulations of terrestrial planet formation from a ring (Hansen, 2009; Kaib and Cowan, 2015; Lykawka, 2020; Nesvorný et al., 2021).

*3.1. Normal disc*

Our results indicate that cases with a normal power law gas disc have difficulties to maintain the radial confinement of the mass distribution that is needed to reproduce the terrestrial planet system.

Figure 4 shows the snapshots of mass versus semi-major axis for a simulation with a normal gas disc with $\tau_{decay}$ = 1 Myr. In this case, an embryo with a few times the lunar mass forms within 0.1 Myr in the Earth region (blue region in the figure; the grey, yellow and red regions are the Mercury, Venus and Mars regions respectively). Compared to the results of the classical model shown in Figure 2 of Woo et al. (2021), the emergence of embryos is much faster in this case, owing to the fact that the initial local solid surface density in the ring model is more than 4 times higher than in the MMSN model.

Due to the rapid formation of lunar to Mars size embryos, these embryos experience strong tidal torque from the gas disc and hence migrate inward through type-I migration (Tanaka et al., 2002; Tanaka and Ward, 2004). After one million years, already 6 Mars size embryos have migrated into the Mercury region (grey region). 4 out of 6 of them are lost to the Sun within 2 Myr through continuous inward Type-I migration.

At this time embryos that form more slowly in the Earth region also start migrating, replenishing the Mercury-Venus region. As the gas disc keeps dissipating, migration slows down and the embryos stop migrating inwards after ~3 Myr. This allows some of the embryos and planetesimals to be scattered outward into the Mars region (red region) towards the end of the gas disc phase (i.e. at about 5 Myr for $\tau_{decay}$ = 1 Myr). At this point, the embryo distribution spans ~1 AU from the Mercury region to the Mars region; accretion of planetesimals continues and embryos slightly grow from 5 to 10 Myr. Giant impacts between embryos are rare during this period: only 3 giant impacts have been recorded in this simulation because the orbits of the giant planets are kept circular during our 10 Myr simulations. We expect that, when the giant planet instability occurs (Gomes et al., 2005; Tsiganis et al., 2005; Morbidelli et al., 2007), the orbits of the embryos would be excited and as shown in previous studies (Clement et al., 2018), giant impacts would occur more frequently (DeSouza et al. 2021).

One noticeable outcome from this normal disc simulation is that the embryos never remain narrowly distributed throughout the whole simulation, due to inward type-I migration and mutual scattering. Embryos formed in this simulation are much more spread than in the initial condition adopted by previous studies in a gas free environment (Hansen, 2009; Kaib and Cowan, 2015; Lykawka, 2020; Nesvorný et al., 2021); in those studies the spread was only a few times 0.1 AU. Hence, the initial conditions adopted in those studies appear questionable, even if the original planetesimals formed in a ring.

Another crucial finding from the normal disc simulation is that the mass-distance distribution of the embryos does not peak at the Venus-Earth region. It is unlikely that the mass distribution of the terrestrial planets, with a massive Earth and Venus at the centre and low-mass Mercury and Mars at the edges, can emerge from such an embryo distribution with a reasonable probability because mutual encounters between the embryos during a later giant impact stage will predominantly spread them apart even farther.

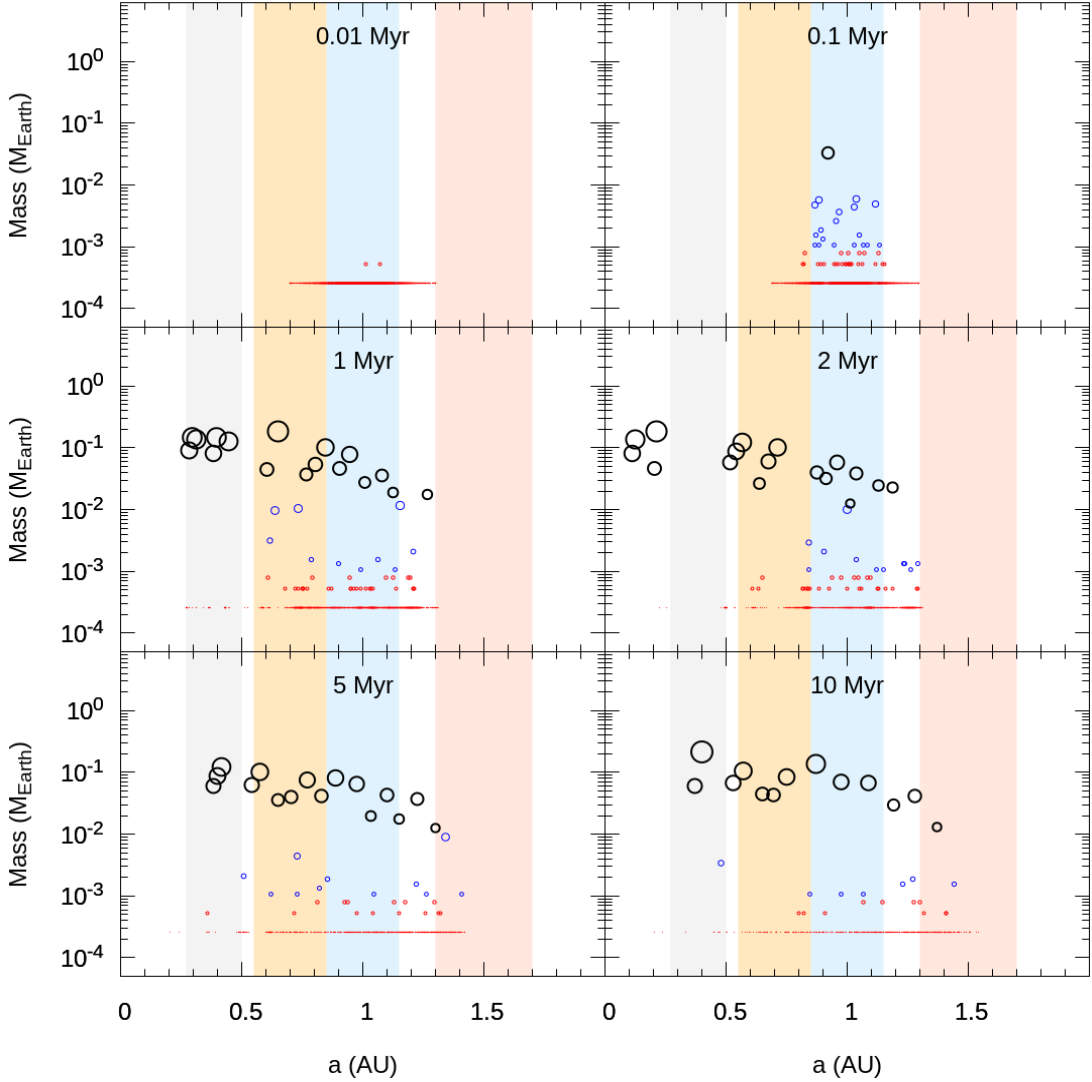

Figure 4 - Snapshots of mass versus semi-major axis for a simulation with a normal gas disc. The gas dissipation timescale, $\tau_{decay}$ = 1 Myr and the mean initial location of the ring of planetesimals, $\mu$ = 1 AU. Thick black circles represent embryos with mass > $10^{-2}$ $M_{Earth}$, blue circles represent intermediate size objects with mass in between smallest planetesimals (red circles) and embryos. All masses are self-gravitating and the colour difference is only for visualisation purposes. Different terrestrial planet regions are shaded (Mercury - grey; Venus - Orange; Earth - blue; Mars - red) according to the definition of Brasser et al. (2016).

Increasing the gas dissipation timescale ($\tau_{decay}$) or the initial central location of the ring ($\mu$) does not help solving the above-mentioned problems. Figure 5 (central and right panels) depicts the mass-distance distribution of embryos at the end of all the simulations we performed for this study adopting a normal power law gas disc and $\tau_{decay}$ = 1 or 2 Myr. As expected, simulations with a longer $\tau_{decay}$ lead to even worse results. Due to the longer lifetime of the gas disc, more embryos are lost due to inward Type-I migration. In simulations with $\tau_{decay}$ = 2 Myr embryos are in general less massive than in simulations with $\tau_{decay}$ = 1 Myr. The mass loss can be > 0.7 $M_{Earth}$ in cases with $\tau_{decay}$ = 2 Myr. Moving the mean initial location of the planetesimal ring from μ = 1 AU to 1.3 AU mitigates the mass loss problem to some extent, as this provides more room for embryos to migrate before being lost into the Sun. Increasing μ, however, cannot solve the lack of mass concentration at Venus-Earth region as radial spreading of the ring materials is still vigorous.

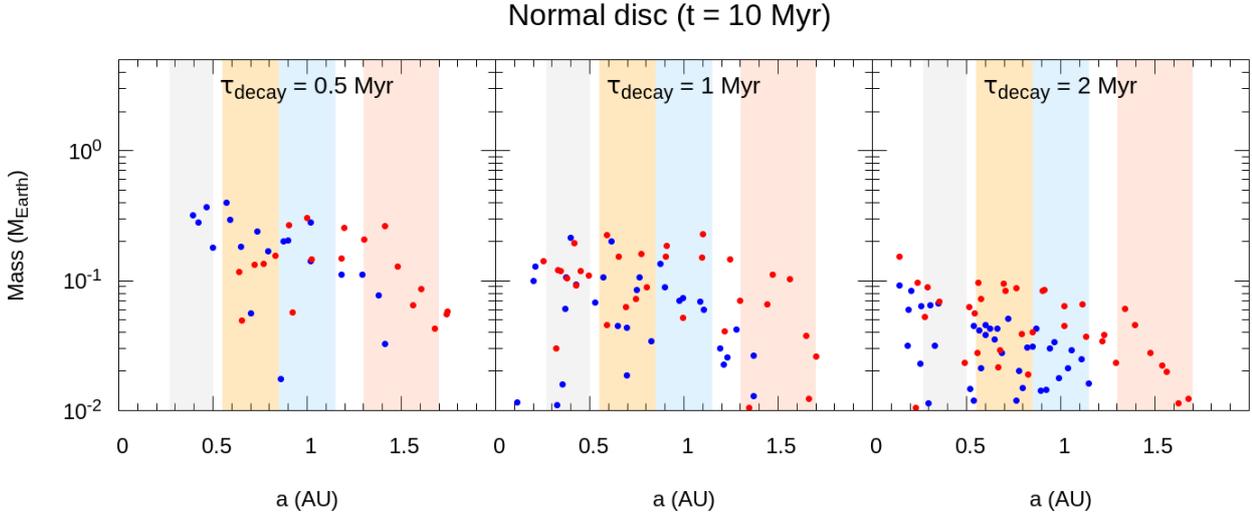

Figure 5 - Masses versus semi-major axis of the embryos at the end of all simulations for the normal gas disc. Blue points are embryos formed in simulations with an initial planetesimal ring at μ = 1 AU, whereas red points are embryos formed in simulations with an initial planetesimal ring at μ = 1.3 AU. Surviving planetesimals are not illustrated for simplicity.

### 3.2. *Beating down migration*

The results illustrated above demonstrate that Type-I migration is a real issue in the ring model, due to the rapid growth of embryos. There are two potential venues to limit the growth of the embryos on the decay timescale of the disk: the first is to reduce $\tau_{decay}$, the second is introducing planetesimals over a time interval, so that initially there is a smaller concentration of mass and embryos growth is slowed down. We explore both avenues below.

#### 3.2.1. Very short disc lifetime

If $\tau_{decay}$ is as short as 0.5 Myr, we see little evidence for a systematic migration of embryos towards the Sun. No embryos reach the inner edge of the disc. Nevertheless, mutual scattering among embryos and with planetesimals still leads to radial spreading of the material in the ring (see Figure 5, left panel). The simulation starting from a ring centred at 1 au also fails to show a peak of embryo mass in the Venus-Earth region; those starting from a ring centred at 1.3 au show such a peak, but rather offset between Earth and Mars.

The results in this case are somewhat reminiscent of those illustrated in Fig. 14 of Walsh and Levison (2016), with embryos ranging from 0.5 to 1.6 au at 10 Myr, which then lead to a good terrestrial planet mass-orbital distribution (Fig. 15 of Walsh and Levison, 2016). But there is a big difference. Because Walsh et al. neglect migration, they can assume a long disk lifetime ($\tau_{decay}$ = 2 Myr) without losing embryos into the intra-Mercury region. Because gas damping keeps the system dynamically cold, the transfer of mass from planetesimals to embryos is much more efficient in the simulation of Walsh and Levison than in ours: at 10 Myr almost all of the mass is in embryos in their simulation, whereas 30% of the mass remains in planetesimals in our case. Consequently, at the end of the disc lifetime their system of embryos is more massive and excites itself dynamically more violently, leading to the formation of planets as massive as Earth and Venus via numerous giant impacts. Instead, in our case, simulations that we have tested on longer timescales – even assuming a sudden transition of the giant planet orbits to the current ones at 10 Myr – show a limited number of giant impacts and fail to produce planets as massive as Earth and Venus, but instead end with a system of numerous, spread-out mini-planets.

Even assuming a very short life time for the gas disc can yield remarkable differences when compared to the gas free simulations that performed by previous study. When compared to the high resolution simulation performed by Morishima (2008), we found that their embryos form extremely rapidly, with 3 embryos reaching ~0.3 to 0.8 $M_{Earth}$ at 10 Myr (their Fig. 2), while in our case none of them growth beyond 0.3 $M_{Earth}$ in the Earth region at 10 Myr. Besides, the three massive embryos in Morishima's simulation spread only ~0.6 AU at 10 Mry (their Fig. 1), similar to the width of their initial ring of planetesimals (0.5 AU), while ours spread to > 4 times of the original width of the ring (defining the full width of our ring as $2\sigma$ = 0.2 AU). This indicates that even with a short life gas disc, it is still sufficient to damp down the orbits of the planetesimals, leading to less crossing of their orbits and thus slower growth of embryos and stronger mutual scattering between them.

### 3.2.2. Introduction of planetesimals over a time interval

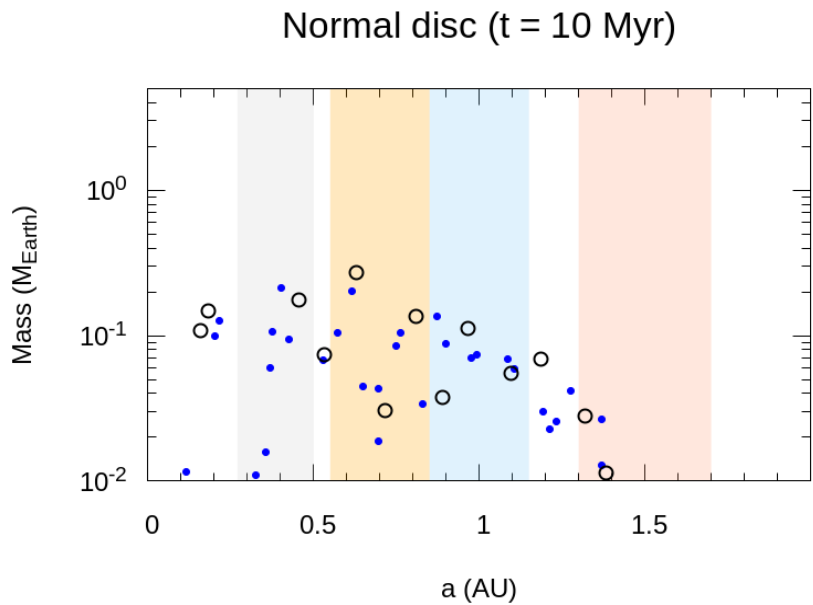

Figure 6 - Same as Figure 5, but only for $\tau_{decay}$ = 1 Myr and $\mu$ = 1 AU. The blue points represent data from simulations with all planetesimals at the beginning (showed in Figure 5), whereas the black hollow circles are data from simulations in which planetesimals are put in gradually within the first 0.5 Myr.

We also introduce planetesimals over the first 0.5 Myr in every 0.1 Myr, in order to slow down the initial growth of the embryos. Figure 5 shows the results of this simulation at 10 Myr.

Compared to the case with all the planetesimals present initially, we found that there are no obvious differences between the two. We still obtain a wide spreaded system with a lack of mass concentration at 1 AU. Hence, we conclude that gradually introducing planetesimals does not mitigate the migration and spreading problem in the normal gas disc.

### 3.3. *Nominal convergent disc*

Given that our attempts to beat-down migration have failed, we modify the gas disc structure to turn migration into our favour, i.e. to concentrate mass radially and form massive embryos in the Venus-Earth region. There is actually no firm constraint on the initial gas surface density of the (inner) solar system. Therefore, we test here the effects of a gas disc with a positive slope surface density within 1 AU, which is the "Nominal convergent disc" set in Table 1. There is some literature precedent for this, because Andrews et al. (2016) report ring substructures around TW Hya down to 1 au.

Figure 7 shows the snapshots of mass versus semi-major axis for a simulation with such a disc. The growth rate of embryos is similar to Figure 4 (MMSN-like disc) within the first 0.1 Myr. Two lunar size embryos are formed within 0.1 Myr. Nevertheless, the growth patterns in the simulations with the normal disc and the convergent disc start to diverge after 1 Myr. Embryos formed within 1 AU migrate outward owing to the positive slope of the gas disc density profile which triggers a positive corotation torque (Masset et al., 2006), while those formed beyond 1 AU would still experience a negative tidal torque due to the negative density slope similar to the case of the normal disc. The convergent migration of embryos helps concentrate the radial mass distribution and create a peak in embryo mass in the Earth region after 1 Myr.

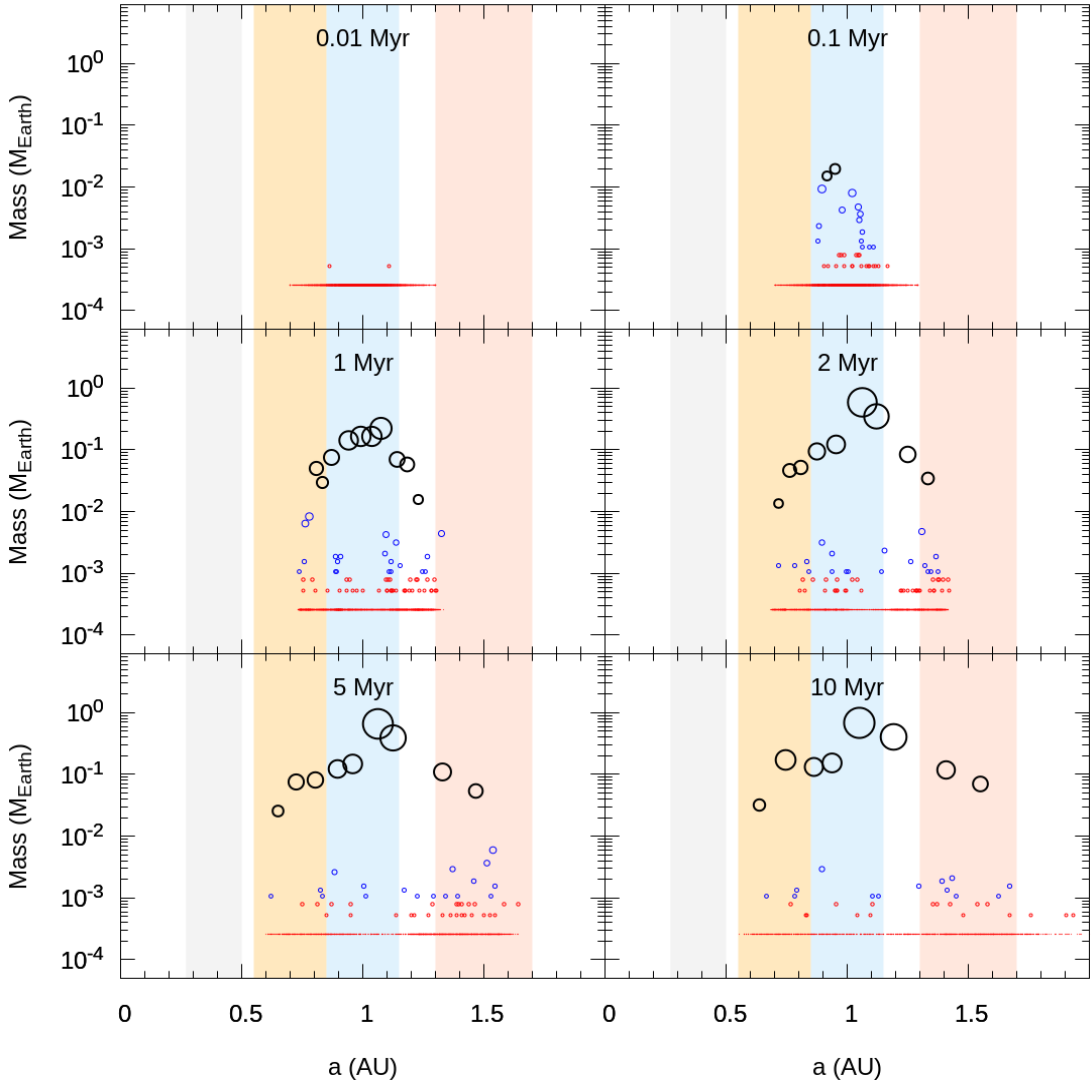

Figure 7 - Same as Figure 4, but for the nominal convergent disc.

The largest embryos in the Earth region grow to 0.5 $M_{Earth}$ after 2 Myr, which is much more massive than in the normal disc case (< 0.1 $M_{Earth}$) at the same time. As the gas disc continues to dissipate, the ring of material radially spreads from the Earth region and embryos enter the Venus region and the Mars region. There are no collisions between embryos in between 2 to 5 Myr due to the radial spreading of the ring. After 5 Myr, the embryos span a region of ~1 AU in width, from the inner Venus region to the middle of the Mars region. At the end of the simulation (10 Myr), the largest embryos in the Earth region reaches ~0.8 $M_{Earth}$, in contrast to only reaching Mars size in the normal gas disc. The Mercury region remains empty throughout the whole simulation. Embryos in the Mars region have approximately the mass of Mars.

Although the results from the convergent disc have a higher potential to match the concentrated mass in the Venus-Earth region of the current terrestrial planet system than those adopting the normal disc, the final embryo distribution is significantly different from the initial conditions adopted by previous studies of the ring model (Hansen, 2009; Kaib and Cowan, 2015; Lykawka, 2020; Nesvorný et al., 2021): it is more spread-out and not uniform in mass. Thus, new simulations of the gas-free phase would be needed. However, we remark that the Earth has almost reached its final mass within 10 Myr, which is difficult to reconcile with the chronology of Earth formation constrained by the Hf-W and U-Pb chronometers (see Sect. 3.5), and that the small total mass in the Mercury-Venus region makes the formation of a Venus analogue unlikely. This suggests that the mass is too concentrated in the Earth region within this set up.

We then tried to introduce the planetesimals over the first 0.5 Myr in every 0.1 Myr in the nominal convergent disc, same as what we did in Section 3.2.2. Figure 8 shows the results of this simulation at 10 Myr. Same as Figure 6, the results between forming all the planetesimals at once and gradually are indistinguishable.

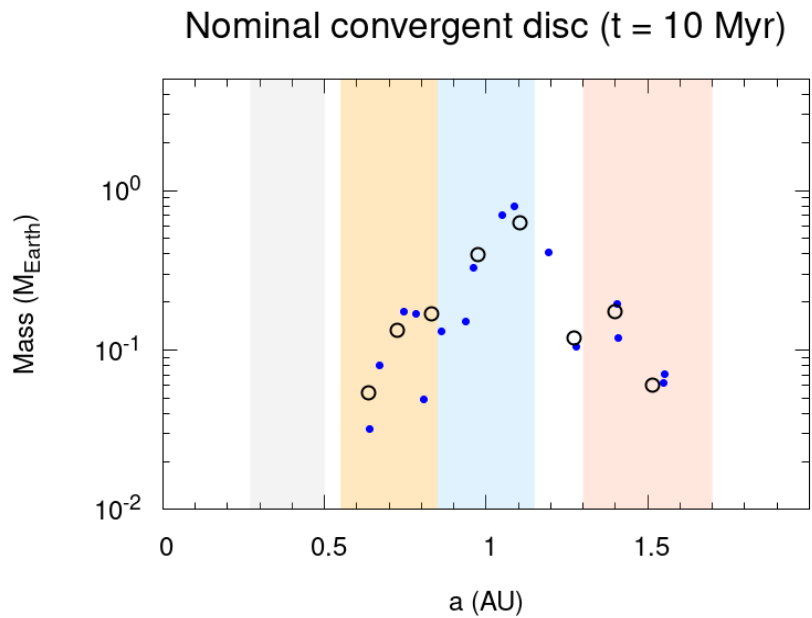

Figure 8 - Same as Figure 6, but for the Nominal convergent disc.

### 3.4. *Convergent shallow inner disc*

The conclusion of the previous section prompts us to adjust the initial gas surface density of the convergent disc within 1 AU to make its radial gradient shallower. This decreases the magnitude of the positive torque experienced by the embryos formed within 1 AU and is expected to lead to a less radially concentrated system. We dubbed this case "Convergent shallow inner disc". Figure 9 depicts the snapshot of mass versus semi-major axis for a simulation with this disc set-up and with the same $\tau_{decay}$ and $\mu$ as for the disc adopted in Figure 7.

The first 1 Myr of the simulation is similar to the non-shallow inner disc. Their differences start emerging after 2 Myr. It is obvious that more lunar to Mars size embryos are formed in the Venus region in the shallow inner disc than in the non-shallow inner disc. The mass of the largest embryos in the Earth region is only ~0.25 $M_{Earth}$, which is more than 3 times less massive than in the steeper inner disc at the same time discussed in the previous subsection. As the positive tidal torque acting on embryos within 1 AU is weaker when the inner disc is shallower, fewer embryos that form in the Venus region end up in the Earth region. Hence more mass remains in the Venus region and thus the overall system is less concentrated.

Towards the end of the gas disc phase (5 Myr), some embryos diffuse into the Mercury region, which does not occur in the steeper inner disc. At the end of the simulation, the embryos monotonically increase in mass from the Mercury to the Earth region. The combined mass of embryos in the Venus region is much higher than in the previous case. The largest mass of embryos in the Earth region is only about 0.4 $M_{Earth}$, less than half of the previous case. A Mars analogue has also formed by 10 Myr.

Overall, this particular case appears more satisfactory. In an upcoming paper we will continue this kind of simulation to ~ 100 Myr, in order to investigate the complete formation of terrestrial planets and the stochastic variability of the results.

Convergent shallow inner disc, $\tau_{decay}$ = 1 Myr, $\mu$ = 1.0 AU

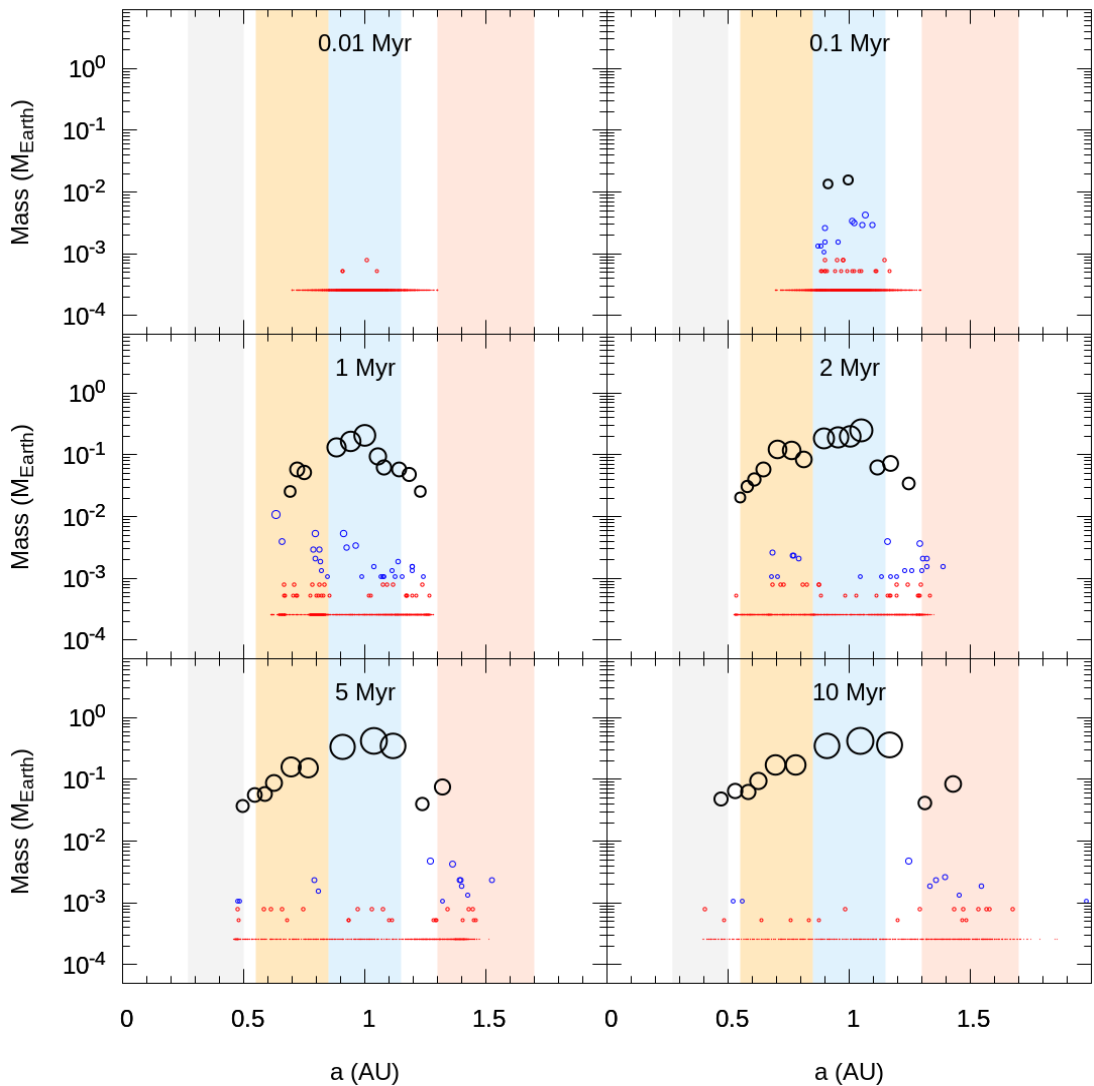

Figure 9 - Same as Figure 7, but for the convergent disc with a shallow radial density gradient in the inner part.

We also tested both shallow and non-shallow convergent discs with a longer $\tau_{decay}$ and a further $\mu$. Figure 10 shows the mass versus semi-major axis for all simulations. Adopting a longer gas disc dissipation timescale results in a sharper peak of the mass distribution in the Earth region in both cases because the gas disc lasts for a longer time, so that more embryos convergently migrate into the Earth region from both the Venus and Mars regions. In the nominal convergent disc, embryos can even grow beyond an Earth mass within 10 Myr when the gas dissipation timescale is long.

Shifting the location of the ring to $\mu = 1.3$ AU has a limited impact on the results, except in the Mars region, where more embryos eventually reside. In some cases, the total mass of the embryos exceeds 3 $M_{Mars}$. This is likely to lead to a final Mars that is too massive, which is the typical outcome of the classical MMSN model (Raymond et al., 2009, 2006).

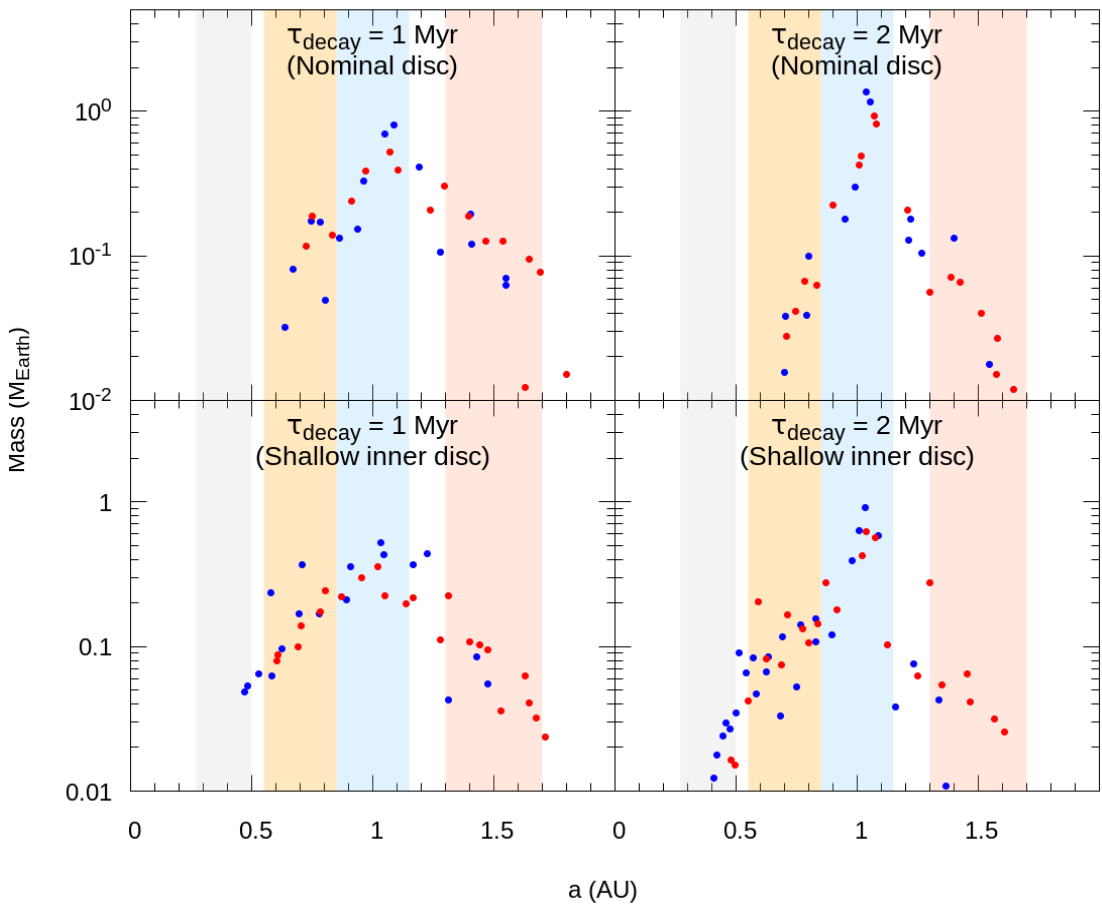

Convergent disc (t = 10 Myr)

Figure 10 - Same as Figure 5, but for the Nominal convergent disc and the Convergent shallow inner disc.

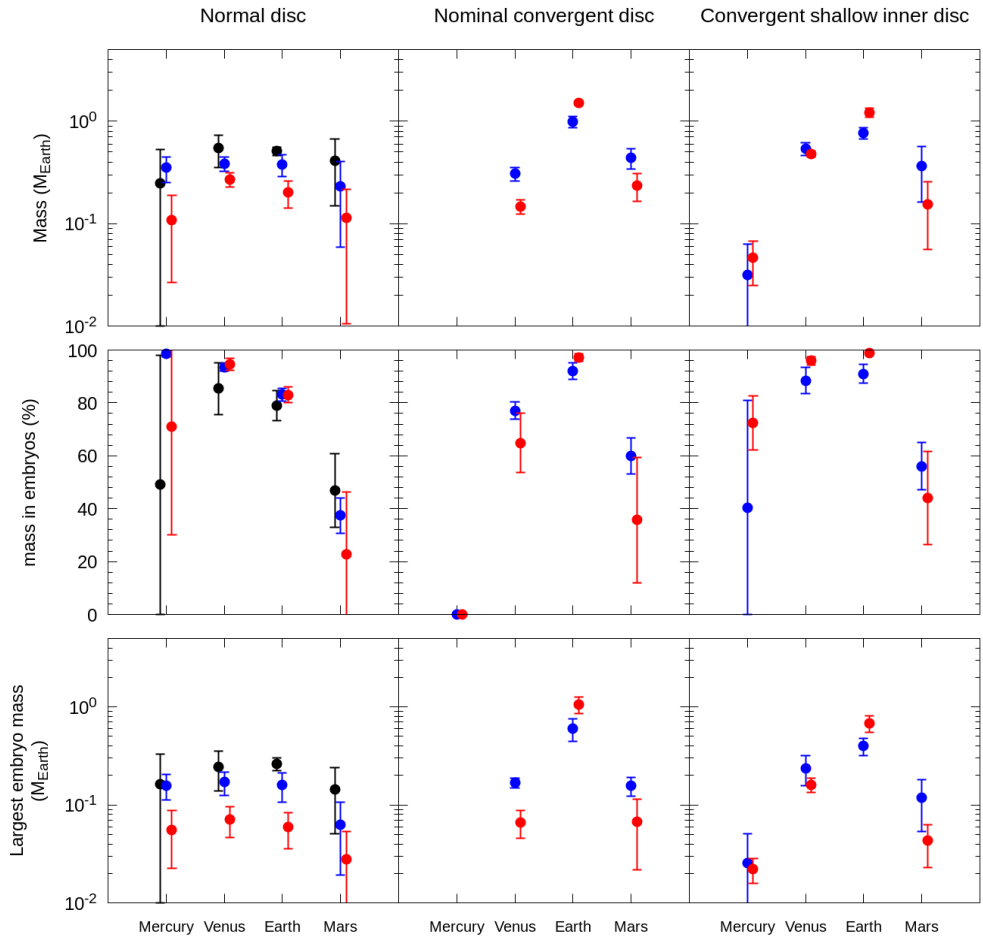

Figure 11 - Mean total mass (planetesimals and embryos), mean percentage of mass in embryos and mean mass of the largest embryos in each planet region at the end of all simulations (10 Myr). The error bars denote 1 standard deviation. Black points (only in the Normal disc) are data from $\tau_{decay}$ = 0.5 Myr, blue points are data from $\tau_{decay}$ = 1 Myr, whereas red points are data from $\tau_{decay}$ = 2 Myr. Results of $\mu$ = 1.0 AU and $\mu$ = 1.3 AU are combined.

### *3.5. Statistical comparison between results obtained with different gas discs*

We made a quantitative comparison among the results obtained in the framework of the normal disc, the nominal convergent disc and the convergent shallow inner disc. Figure 11 shows the mean total mass (planetesimals and embryos), mean percentage of mass in embryos (as opposed to planetesimals) and the mean mass of the largest embryo in each planet region for each simulation set. It is obvious that the total mass of the system is distributed rather evenly in the normal disc case, while the total mass of the system peaks in the Earth region in the convergent disc cases, consistent with the distribution of embryo's mass observed in Figure 5 and figure 10. The same also applies for the mean largest embryo mass. This implies that the examples shown in Figure 4, Figure 7 and Figure 9 represent the general situation.

The statistical values for the Mercury region have the largest uncertainties. The Mercury region is entirely empty initially, so its final total mass depends on how many embryos migrate or scatter into this region, which has a significant stochastic variability. The same also applies for the Mars region, but to a lesser extent as in some cases there are initially planetesimals in the Mars region (e.g. $\mu$ = 1.3 AU).

As pointed out in the previous section, results from the normal disc generally do not reproduce the concentration of mass in the Venus-Earth region of today's system, so we will focus below on comparing data between the two convergent disc cases in each planet region.

The Mercury region is empty in the nominal convergent disc at the end, while a substantial mass of the same order of magnitude as the real Mercury mass could enter the Mercury region when a shallow inner disc was adopted. Planetesimals can also be scattered into the Mercury region in a shallow inner disc, as highlighted by a non-trivial percentage of mass made by planetesimals in that region. A shallow inner disc is thus preferred in order to promote the formation of Mercury.

The difference between the total mass in the Earth region and the Venus region is the most crucial criterion in determining which convergent disc works better. This can be quantified by the mass ratio of the Earth region to the Venus region ($M_{Earth}/M_{Venus}$). According to Figure 11, this ratio is always > 3 in the nominal convergent disc case, while it decreases to < 1.2 with a shallow inner disc, similar to the current value 1.23. The shallower inner disc allows more mass to stay in the Venus region, as explained in the previous section.

The current large $M_{Earth}/M_{Mars}$ is best reproduced by $\tau_{decay}$ = 2 Myr, as more embryos migrate toward 1 AU. However, Earth tends to grow too fast when $\tau_{decay}$ = 2 Myr (see Section 3.5). Thus there appears to be a delicate trade-off between the two regions.

Since the Venus-Earth region is the densest region in terms of solid mass, these regions form embryos rapidly and so typically most of their mass (> 70 %) at 10 Myr is in the form of embryos. However, this is not the case in the Mars region. Only about 30 to 60% of the mass in the Mars region is comprised in embryos. Given that the largest embryo in the Mars region is already ~1 $M_{Mars}$, the large remnant planetesimal mass in the Mars region may lead to an overgrowth of Mars. This potential problem will need to be assessed with longer simulations.

Overall, the results obtained in the simulation with a convergent shallow inner gas disc and $\tau_{decay}$ = 1 Myr appear to have the highest potential to reproduce the real terrestrial planets in the end. We will test this hypothesis thoroughly in a forthcoming manuscript.

3.6. *Comparison with constraints on Earth's and Mars' formation timescales*

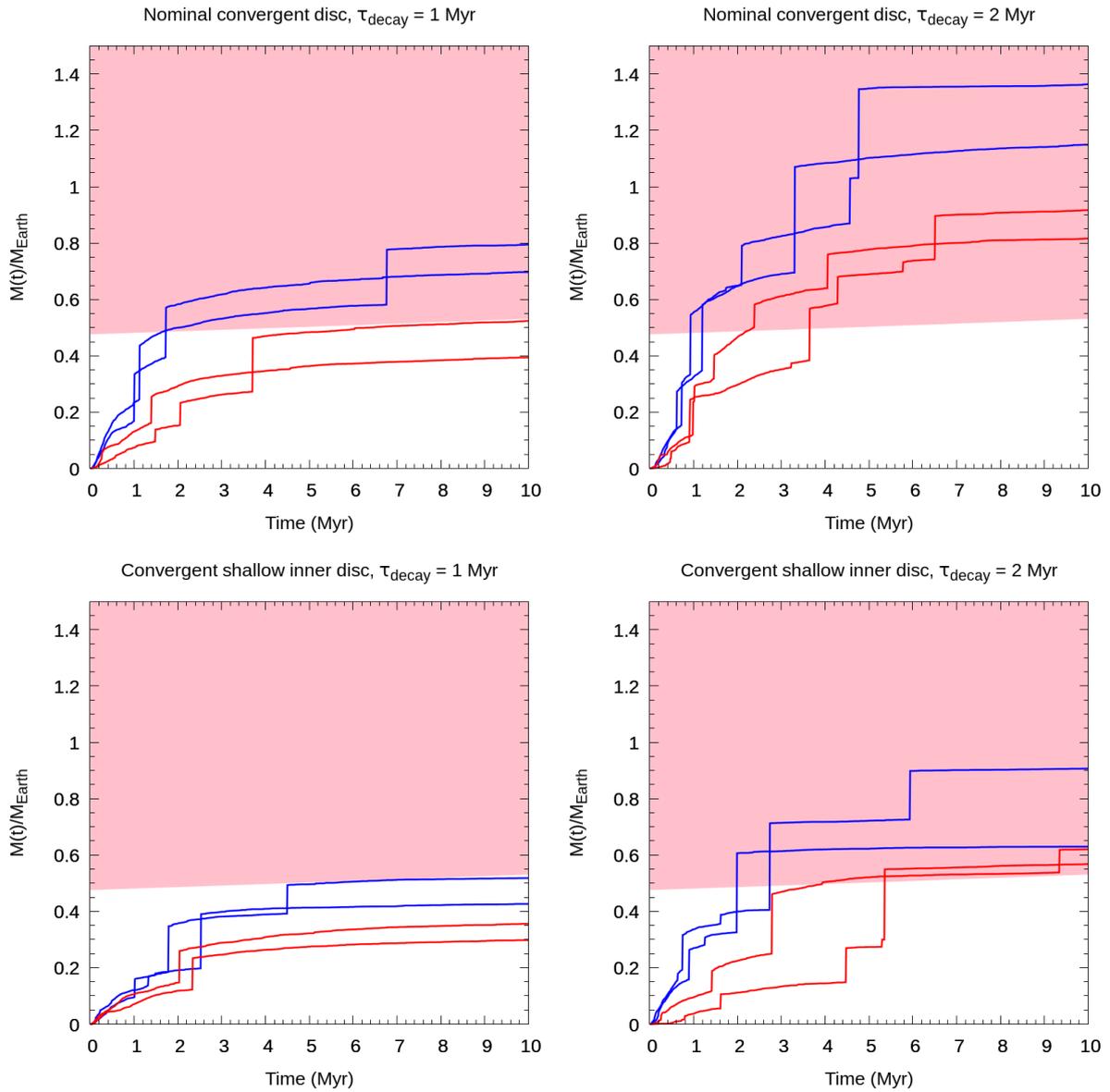

Figure 12 - Mass evolution of the most massive embryos in the Earth region at the end of each simulation with a convergent disc (nominal and shallow inner disc). Blue lines are simulations starting with a solid ring at $\mu = 1$ AU, whereas red lines are simulations starting with a solid ring at $\mu = 1.3$ AU. The shaded pink region indicates the forbidden region referred from the Hf-W data of Earth's samples (Rudge et al., 2010), assuming a metal-silicate equilibration of $k = 0.4$.

The Hf-W and U-Pb systems provide a constraint on the chronology of an object's metal-silicate differentiation. Its interpretation in terms of planet growth history depends on a number of assumptions: continuous or episodic core formation, and the degree of metal-silicate equilibration, $k$, after each accretion event (particularly giant impacts from other embryos). By considering an exponential growth for Earth of type $M(t) = M_{Earth}(1 - e^{-t/\tau})$, Rudge et al. (2010) showed that the characteristic timescale for Earth formation $\tau$ provided by the Hf-W chronometer is consistent with that provided by the U-Pb chronometer if the degree of metal-silicate equilibration during accretion is $k = 0.4$. Using this value, Rudge et al., could provide broad bounds to Earth accretion history. In the 0-10 Myr interval, this amounts to excluding that the Earth exceeded 0.5-0.6 Earth masses. This boundary is illustrated by the shaded area in Figure 12. In the same figure, the various plots show the mass evolution of the most massive embryos in the Earth's region at the end of each simulation. The most massive embryos have fastest growth in the nominal convergent disc with $\tau_{decay} = 2$ Myr and, in general, the most massive embryos in the Earth region grow faster when the initial ring is at 1 AU. As one can see, only the simulations conducted in the framework of the convergent shallow inner disc are consistent with the upper bound on Earth's growth of Rudge et al. (2010), marginally for all values of $\mu$ if $\tau_{decay} = 1$ Myr, but only if $\mu = 1.3$ AU if $\tau_{decay} = 2$ Myr. Instead, the simulations conducted with the nominal convergent disc are not consistent with this bound, except in the case $\tau_{decay} = 1$ Myr, $\mu = 1.3$ AU.

A more accurate test against the Hf-W chronometer constraint requires to evaluate, at each accretion episode in our simulation, the degree of metal/silicate equilibration and the partition coefficient of W, depending on the impactor's mass, the Earth's pressure and temperature at the core/mantle boundary and the mantle's oxygen fugacity (e.g. Jennings et al., 2021). This will be the object of future work, once the simulations are brought to completion. For the moment, we use Figure 12 as an indication that terrestrial planet formation from a ring of planetesimals is much more likely to be successful dynamically in a disc with weak convergent migration (as our convergent shallow inner disc) than in a disc with strong convergent migration (as in our nominal convergent disc). We remark that the "successful" simulations of Brož et al. (2021) (using the nominal convergent disc) all seem to grow the Earth too fast with respect to the bounds provided by Rudge et al. (2010).

On the contrary of Earth, the time constraint on Mars' formation is believed to be more firm, as Mars suffers far fewer giant impacts compared to Earth and thus is expected to experience a high degree of metal/silicate equilibration during accretional impacts. This is supported by our simulation results. Figure 13 shows the growth curve of the largest embryos in the Mars region in the simulations comprising convergent discs. Compared to Earth's growth, Mars' growth is much more gradual, with much fewer big jumps, indicating that Mars' growth is mainly a result of planetesimal accretion instead of giant impacts. Assuming growth via planetesimal accretion and full equilibration, the constraints of the Hf-W chronometer on Mars' accretion history have been elucidated by Dauphas and Pourmand (2011) and are illustrated by the black curve and yellow-green shaded area in Figure 13. However, other results suggest that the role of giant impacts should have been non-negligible for Mars as well and argue for a protracted accretion of Mars (Zhang et al., 2021).

Different from Earth's situation, a longer $\tau_{decay}$ results in a smaller and more slowly growing Mars in general because of the migration issues already discussed. Although forming Mars analogue that is too massive can happen when $\tau_{decay} = 1$, there are also very close matches with the growth timescale suggested by Dauphas and Pourmand (2011) in both convergent discs, particularly if the initial ring is located at $\mu = 1.3$ AU. Instead, simulations with $\tau_{decay} = 2$ Myr systematically form a Mars mass planet too slowly and still too small after 10 Myr.

To summarise, combining the Earth and Mars cases, the results suggest that a convergent shallow inner disc with a short $\tau_{decay}$ (~1 Myr) is the most promising to produce terrestrial planets consistent with their mass, orbit and chronological constraints.

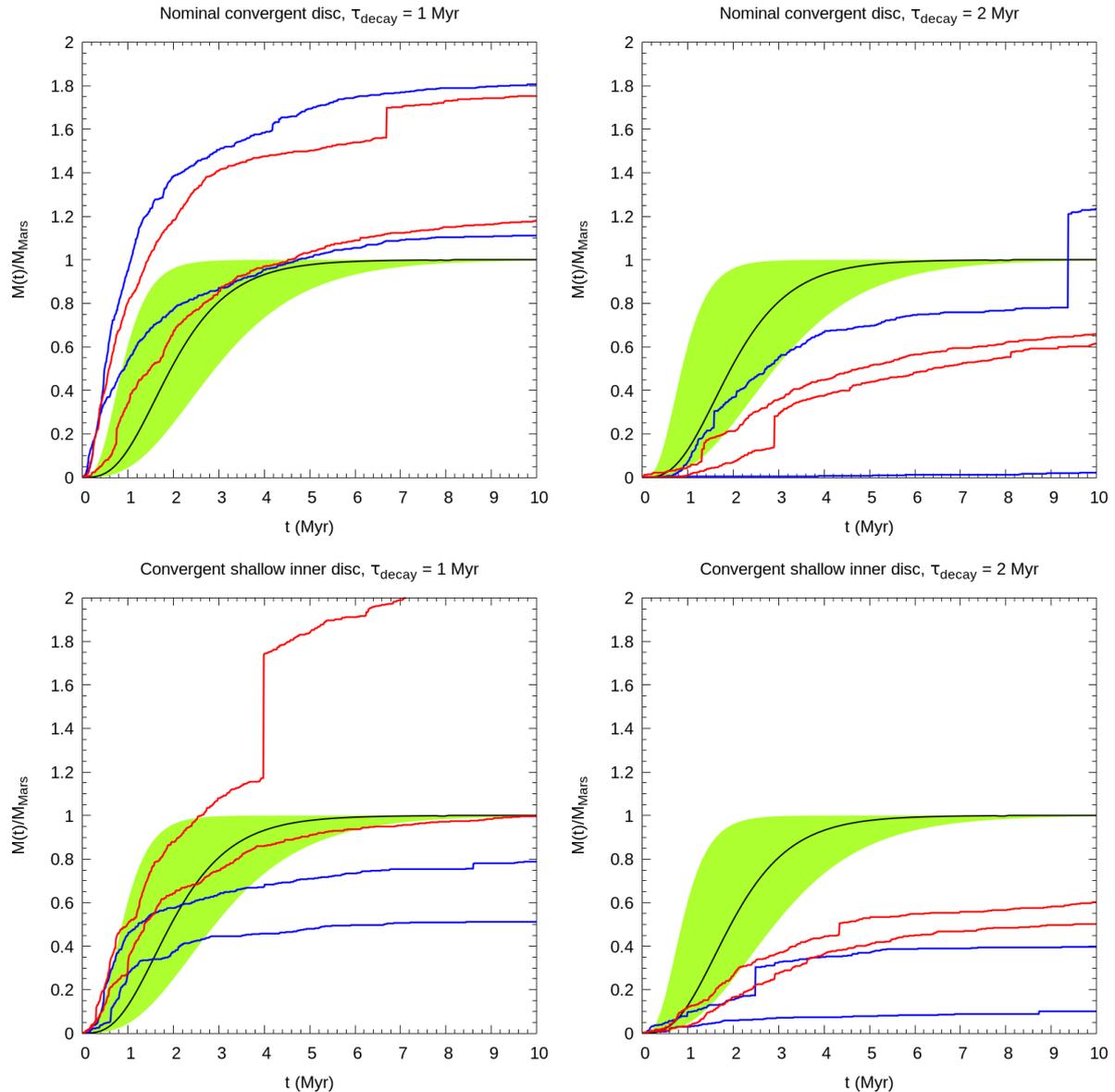

Figure 13 - Same as Figure 12, but for the most massive embryos in the Mars region. The black line is the growth curve suggested by Dauphas and Pourmand (2011) based on the Hf-W chronology on Shergottites. The yellow-green band represents the 95% con-fidence interval of their model.

## 4. Discussion

We demonstrated that it is possible for the ring model to match the mass distribution and isotopic chronological constraints, preferentially from a convergent gas disc with a shallow inner region. However, it does not mean that the ring model is perfect in all aspects to represent the solar system's history. There are also some potential uncertainties for completing the full story of the terrestrial planet formation within the context of the ring model.

### 4.1. Dynamical perspective

Our simulations have run for only 10 Myr (the gas-disc phase). Because our embryo mass-distance distribution is different from those assumed in previous studies of the giant impact phase (i.e. Hansen, 2009; Kaib and Cowan, 2015; Lykawka, 2020; Nesvorný et al., 2021), we cannot rely on those simulations to claim a successful reproduction of the terrestrial planets of the solar system. Longer simulations extending to > 100 Myr will have to include a phase of giant planet instability during which the orbits of the giant planets evolved from quasi-circular and coplanar orbits to the current, moderately excited ones, sometimes while the terrestrial planets were still forming. We will present long-term simulations in a forthcoming manuscript.

A specific attention will have to be paid to planet Mercury, which is particularly difficult to reproduce in terms of final mass, semi-major axis ratio with Venus and core/mantle mass ratio (Clement et al., 2019, 2021; Clement and Chambers, 2021; Lykawka and Ito, 2017, 2019). One idea is that Mercury is the lone survivor of a series of erosive collisions of inner embryos (Clement et al., 2021). In our convergent shallow inner disc simulation, we do not form more than 1 embryo per run in the Mercury region. Hence, this explanation could not apply to our Mercury candidate. However, we can envision that the starting planetesimals were differentiated (due to their early formation) and that those near the inner edge of the ring lost part of their mantle in erosive collisions (Hyodo et al., 2021). If this is true (to be modelled in a future paper) then Mercury can naturally form with a high Fe/Si ratio by accretion of these eroded planetesimals.

The asteroid belt is assumed to be empty in the simulations presented here, but clearly asteroids exist in this region, even if with an anecdotal small mass. It is possible that some planetesimals formed in the asteroid belt region, possibly towards the end of the gas-disc lifetime, consistent with the chronology of accretion of chondritic parent bodies. In absence of resident embryos, their dynamical excitation would occur only during the giant planet's dynamical instability, sometime after the removal of gas from the disc (Deienno et al., 2018). It has also been suggested that all non-carbonaceous asteroids formed in the terrestrial planet ring and have been scattered into the asteroid belt during terrestrial planet growth (Raymond and Izidoro, 2017a), while it is now widely accepted that all carbonaceous asteroids have been implanted during the growth and migration of the giant planets (Raymond and Izidoro, 2017b; Walsh et al., 2011).

### 4.2. Isotopic perspective

Perhaps a bigger challenge of the ring model arises from the chemical and isotopic perspective. The chemistry of the Earth's mantle, particularly the partition of moderately siderophile elements and the depletion pattern of volatile elements, suggests that the Earth accreted from a diversity of materials, characterised by different oxygen fugacity (Rubie et al., 2011, 2015) and different volatile depletion (Sossi et al., 2022). Also, Mars, Earth and the non-carbonaceous meteorites (both chondrites and achondrites) show a significant spread in nucleosynthetic isotopic anomalies (e.g. $^{48}$Ca, $^{54}$Cr, $^{50}$Ti, $^{92}$Mo, $^{100}$Ru, $^{96}$Zn, etc) (e.g. Burkhardt et al., 2021; Fischer-Gödde and Kleine, 2017; Render et al., 2017; Trinquier et al., 2009, 2007; Warren, 2011; Yamakawa et al., 2010). How could such a variety of materials reside in a compact ring?

Clearly some modification on the ring model will be required in order to explain these heterogeneities. It may be possible that a low-mass asteroid belt, of distinct chemical and isotopic composition from the early planetesimals formed in the ring, provides enough contamination to the terrestrial planets to explain Earth heterogeneous accretion and its differences with respect to Mars. This hypothesis will need to be tested with numerical simulations combined with a chemical evolution model for the Earth (Rubie et al., 2011, 2015; Jennings et al., 2021).

## 5. Conclusions

We revisited the idea of forming terrestrial planets from a ring of planetesimals. Previous works (Deienno et al., 2019; Hansen, 2009; Lykawka, 2020; Morishima et al., 2008; Nesvorný et al., 2021; Walsh and Levison, 2016) suggest that this initial condition leads to the correct mass-distance distribution of the terrestrial planets. Moreover, recent advances on planetesimals formation suggest that the first planetesimals form in rings (Drążkowska et al., 2016; Drążkowska and Alibert, 2017) and, in the inner solar system, such a ring could be around the silicate sublimation line, located close to 1 AU (Morbidelli et al., 2022). We performed 28 high-resolution N-body simulation with GENGA (Grimm and Stadel, 2014; Grimm et al., 2022) of a ring of self-interacting planetesimals, with no arbitrarily introduced planetary embryos, and observed whether the system remained radially confined at the end of the gas-disc phase (10 Myr). We found that, by adopting a normal gas disc with a gas surface density that follows a negative power law similar to the minimum mass solar nebula, Type-I migration acting on the growing planetary embryos, combined with the mutual scattering of embryos and planetesimals, cause the ring of material to spread radially beyond into a 1 AU-wide region. Some embryos also reach the intra-Mercury region by migration. The resulting embryo distribution thus does not resemble that assumed in the initial condition adopted by previous studies of terrestrial planet formation from a ring in a gas-free environment (Hansen, 2009; Nesvorný et al., 2021). We have tried to reduce the effect of migration by assuming a very fast decay of the disc's gas density or introducing planetesimals over a 0.5 Myr-long interval, but in both cases we obtained results that are inconsistent with the current distribution of the terrestrial planets

We thus proposed that the initial gas disc could have a surface density with a maximum at ~ 1 AU (Convergent disc) (Brož et al., 2021; Haghighipour and Boss, 2003; Lyra et al., 2010; Masset et al., 2006; Ogihara et al., 2018; Zhang et al., 2014). In such a gas distribution embryo migration is outward within 1 au and inward beyond 1 au. This convergent migration prevents the radial spreading of the mass distribution, with the most massive embryos forming in the Earth region. The embryo system formed in this process is thus more likely to lead to the concentrated mass-distance distribution of the current system at the Venus-Earth region if the simulation is extended to ~100 Myr timescales. However, if the convergent migration is too strong, as in the nominal disc proposed by Brož et al. (2021), Earth grows too quickly compared to constraints derived from the Hf-W and U-Pb chronometers (Rudge et al., 2010) and there is a too big imbalance between the masses of Earth and Venus. These problems are alleviated if migration is weaker, e.g. if the positive radial density gradient of the gas in the inner disc is shallower (Convergent shallow inner disc). Overall, we found that the Convergent shallow inner disc with a gas disc decay time scale of ~1 Myr leads to a better match with Earth and Mars growth timescales (Dauphas and Pourmand, 2011; Rudge et al., 2010), and potentially as well as the mass-distance distribution of the current terrestrial system.

In the future, we will extend these most promising simulations to timescales of ~100 Myr, also including a phase of giant planet instability. We will also investigate the role of a low-mass extension of the ring into the asteroid belt, potentially contaminating the terrestrial planets and affecting their chemical and isotopic composition.

**Acknowledgement**

This work was funded by the Deutsche Forschungsgemeinschaft (SFB-TRR 170, subproject B6, project no. 176) and the ERC project N. 101019380 "HolyEarth". This work was granted access to the HPC resources of IDRIS under the allocation 2022-A0120413416 made by GENCI. The authors also acknowledge the computational support from Service and Support for Science IT (S3IT) of University of Zurich. JS acknowledges the financial support from the National Center of Competence in Research PlanetS, supported by the Swiss National Science Foundation.